\documentclass[prl,aps,twocolumn,superscriptaddress]{revtex4-1}
\usepackage{amsmath}
\usepackage{latexsym}
\usepackage[english]{babel}
\usepackage{graphicx}
\usepackage{bm}
\usepackage{array}
\usepackage{amssymb}
\usepackage{amsfonts}
\usepackage{multirow}
\usepackage{epstopdf}
\usepackage{color}

\begin{document}

\title{Theory of the magnetic domains phases in ferromagnetic superconductors}

\author{Zh. Devizorova}
\affiliation{Moscow Institute of Physics and Technology, 141700 Dolgoprudny, Russia}
\affiliation{Kotelnikov Institute of Radio-engineering and Electronics RAS, 125009
Moscow, Russia}
\author{S. Mironov}
\affiliation{Institute for Physics of Microstructures, Russian Academy of Sciences,
603950 Nizhny Novgorod, GSP-105, Russia}
\author{A. Buzdin}
\affiliation{University Bordeaux, LOMA UMR-CNRS 5798, F-33405 Talence Cedex, France}
\affiliation{Sechenov First Moscow State Medical University, Moscow, 119991, Russia}

\begin{abstract}
Recently discovered superconducting P-doped EuFe$_2$As$_2$ compounds reveal the situation when the superconducting critical temperature substantially exceeds the ferromagnetic transition temperature. The main mechanism of the interplay between magnetism and superconductivity occurs to be an electromagnetic one and a short period magnetic domain structure was observed just below Curie temperature [Stolyarov et al., Sci. Adv. \textbf{4}, eaat1061 (2018)]. We elaborate a theory of such transition and demonstrate how the initial sinusoidal magnetic structure gradually transforms into a soliton-like domain one. Further cooling may trigger a first-order transition from the short-period domain Meissner phase to the self-induced ferromagnetic vortex state and we calculate the parameters of this transition. The size of the domains in the vortex state is basically the same as in the normal ferromagnet, but with the domain walls which should generate the set of vortices perpendicular to the vortices in the domains.
\end{abstract}

\maketitle

The coexistence of magnetism and singlet
superconductivity has always been of great interest because of their
competing nature. Already V.~Ginzburg \cite{Ginzburg}
showed that uniform magnetism in bulk systems may destroy
superconductivity due to the electromagnetic (EM) mechanism (so-called, orbital effect), i.e.
generation of the screening Meissner currents. In addition, the exchange
field tends to align electron spins parallel to each
other which prevents the formation of Cooper pairs with the opposite spin
directions [exchange (EX) mechanism] \cite{Matthias}. As a result, the
coexistence of {\it uniform} ferromagnetism and superconductivity becomes possible
primary in thin-film structures with the damped orbital effect \cite{Ginzburg}, spin-triplet uranium-based superconductors \cite{Aoki} or
 artificial superconductor-ferromagnet hybrids \cite{Efetov_RMP, Buzdin_RMP, Aladyshkin}.

In contrast, {\it non-uniform} magnetic states may  peacefully coexist with the superconducting ordering. The
typical example is the antiferromagnetic superconductors RRh$_{4}$B$_{4}$
and RMo$_{6}$S$_{8}$ with the rare-earth element R \cite{Maple} where the
net magnetic moment at the scale of the superconducting coherence length $\xi$ is
zero and, thus, does not influence Cooper pairs. Somewhat similar 
cryptoferromagnetic phases were predicted for the ferromagnetic superconductors (FSs) \cite{Anderson} 
%In this case the spatial period of magnetic structure should simultaneously exceed the interatomic distance and be much smaller than the size of a Cooper pair. 
and was lately observed  in ErRh$_{4}$B$_{4}$ \cite{Moncton} and HoMo$_{6}$S$_{8}$ \cite{Lynn} together with the reentrant
superconductivity (see, e.g., \cite{Buzdin} for review).

The early theories of non-uniform magnetism in
FSs accounted only the EM interaction \cite{Krey}. For the isotropic compounds EM effect favors the spiral magnetic texture in the superconducting phase instead of the ferromagnetism \cite{Ferrell, Matsumoto} while magnetic anisotropy
should trigger the formation of domain structures (DS) \cite{Krey, Creenside} which can coexist with Abrikosov vortices 
\cite{Tachiki, Ishikawa, Laiho}. However, the further investigation of these intriguing phenomena in FSs with purely EM interaction was interrupted because it turned out that even small exchange field producing negligible contribution to the magnetic energy should dramatically affect the magnetic texture of FSs \cite{Buzdin}. At the late 80-s there were no FSs where EM interaction could dominate, and the  research became mainly focused on the EX mechanism \cite{Anderson, Fulde, Bulaevskii1, Bul1, Bul2, Bul3}. In principle, the effects of EM interaction could be observed in triplet FSs. However, in all three known triplet FSs \cite{Aoki} the Curie temperature $\theta$ is well above the superconducting critical temperature $T_c$ so that below $T_c$ the magnetic structure is already frozen and insensitive to the superconductivity.

%%%%%%%%
\begin{figure*}[t!]
	\includegraphics[width=1\linewidth]{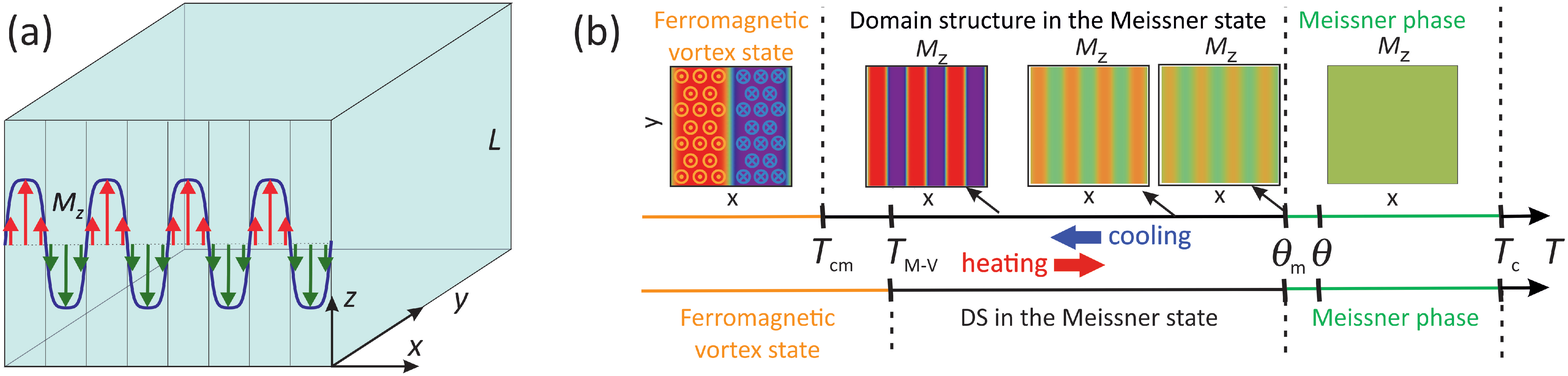}
	\caption{(a) Ferromagnetic superconductor (FS) with the easy-axis magnetic anisotropy along the $z$-axis. (b) The phase diagram demonstrating the temperature evolution of the coexisting phases in FS with dominant EM mechanism.} \label{evol}
\end{figure*}
%%%%%%%%

Recently the interest to the FSs with purely EM
interaction has been unexpectedly renewed with the discovery of P-doped EuFe$_{2}$As$_{2}$ compound where $\theta<T_c$ and superconductivity coexists with ferromagnetism in a broad temperature interval \cite{Ahmed, Cao, Nowik, Zapf1, Zapf2, Nandi}. The rather large critical temperature $T_c \sim$ 20-30 K in the ferroarcenide family \cite{Johnston} and the robustness of superconductivity towards a disorder strongly support the $s$-wave character of the superconducting pairing. In such a case, the exchange field $h_{ex}$ generated by Eu atoms in the low temperature ferromagnetic phase should be rather weak. The upper critical field $H_{c2}(T)$ in EuFe$_{2}$As$_{2}$ compounds is characterized by a large slope at $T_c$:  $dH_{c2}/dT \sim$ 3 T/K \cite{Tsvyashchenko}, which corresponds to the small superconducting coherence length $\xi \sim$ 1.5 nm. A very strong polarization of Eu subsystem is achieved at fields $\gtrsim$ 1 T \cite{Nandi} but it does not  result in any observable decrease of the transition temperature \cite{Tsvyashchenko} and we may conclude that $h_{ex} \ll T_c$ (hereinafter we put the Boltzman constant $k_B=1$). In addition, the time resolved  magneto-optical measurements in EuFe$_{2}$As$_{2}$ \cite{Pogrebna} reveal a very slow relaxation time for Eu$^{+2}$ spin $\tau \sim$ 100 ps, which imply the exchange interaction $h_{ex} \sim \hbar / \tau \sim$ 0.1 K. Moreover, the M\''{o}ssbauer studies \cite{Nowik, Nowik2} and density-functional band-structure calculations \cite{Jeevan} indicate that the exchange interaction between Eu atoms and superconducting electrons in EuFe$_{2}$As$_{2}$ and similar compounds is very weak $h_{ex} \lesssim$ 1 K  and then the exchange RKKY contribution into the magnetic energy $\theta_{ex} \sim N(E_F) h_{ex}^2 \sim$ 10$^{-3}$ K  (for the electron density of states $N(E_F) \sim$ 2-3 states/eV per one Eu atom \cite{Johnston}). 

The strong spin-orbit scattering in EuFe$_{2}$As$_{2}$ is likely to suppress the paramagnetic mechanism of superconductivity destruction. When the spin-orbit electron scattering mean free path $l_{so}$ is of the order of the ordinary mean free path $l$ ($l_{so} \gtrsim l$) the EM interaction dominates over the exchange one in the non-uniform magnetic structure formation, if $\theta_{ex} < \theta (a \xi^2/l^3)$ , where $a$ is of the order of interatomic distance \cite{Bulaevskii}. The small value of the superconducting coherence length in EuFe$_{2}$As$_{2}$ ensures the domination of the EM mechanism.

The unusual relation $\theta<T_c$ in EuFe$_{2}$As$_{2}$ provides an access to the almost unexplored situation when the ferromagnetism nucleates in fully developed superconducting state. The recent pioneering experiments 
on the high-resolution
visualization of the magnetic texture in EuFe$_{2}$As$_{2}$ \cite{Veshchunov, Vinnikov} provide the first direct evidence of the transitions from the short-period domain Meissner state to the phase where magnetic domains coexist with Abrikosov vortices. Interestingly, these transitions reveal hysteresis behavior when varying the temperature \cite{Vinnikov}. However, despite the rapid experimental progress the theory of the magnetic states evolution in anisotropic FSs with purely EM interaction is still lacking.

In this Letter we present the theory of the magnetic domain phases in FSs with low Curie temperature and purely EM interaction. We demonstrate how magnetic domains
evolve from sinusoidal profile to the step-like structures when cooling the sample. Also we calculate the key parameters of first-order transition to the phase with coexisting domains and vortices and suggest the explanation for the hysteresis behavior of the domain structure in EuFe$_{2}$As$_{2}$ \cite{Veshchunov, Vinnikov}. Finally, we show that the domain walls favor the generation of  unusual vortices with the cores perpendicular to the vortices in the domains.

Before going into details we briefly overview possible regimes in the evolution of magnetic texture with the variation of temperature $T$ (see Fig.~\ref{evol}). In the cooling process at $T=T_c$ the superconducting Meissner phase appears. The well-developed superconductivity prevents the nucleation of uniform ferromagnetism at $T=\theta$. As a result, the magnetic order emerges at the temperature $\theta_m<\theta$ and the magnetization has the sine profile with only one spatial harmonic. While cooling below $\theta_m$ the nonlinear effects give rise to other spatial harmonics and the magnetization evolves towards the well-developed step-like DS with increasing domain size. At temperature $T_{M-V}$ the growing amplitude of the magnetization makes the uniform superconducting phase less favorable than the phase with coexisting DS and vortex lattice. However in the cooling regime the immediate vortex entry at $T=T_{M-V}$ is prevented by the Bean-Livingston like barrier. This results in overcooling of the Meissner state and the first order phase transition to the vortex state (VS) occurs only at $T=T_{cm}<T_{M-V}$ when the barrier vanishes. At the same time, in the heating regime the system stays in the VS until $T=T_{M-V}$, thus, demonstrating the hysteresis behavior.

As it is well known in a ferromagnetic thin film sample with perpendicular  anisotropy the domain structure appears in order to minimize the stray  field.  The period of such structure depends on the  thickness of the sample and usually exceeds micron size and it is much larger than that in the domain Meissner phase. Note that the formation of the short period domain Meissner state is related with the volume effect of the interaction between magnetism and superconductivity, while the existence of the domains in normal ferromagnets is related with its demagnetization factor (shape effect). In the considered case these ferrromagnetic domains will be in the vortex state.

To support the above qualitative picture we calculate the temperature evolution of the magnetic texture in the FS with $\theta<T_c$ using the Ginzburg-Landau approach. The free energy functional describing the FS with the strong easy-axis anisotropy along the $z$ axis reads \cite{Buzdin}:

\begin{multline}
\label{F_FS}
F(B_z,M_z)=\frac{(B_z-4\pi M_z)^2}{8\pi}+\frac{A^2}{8\pi \lambda^2}+\\+\frac{n {\tilde \theta}}{M_0^2}\left[\tau M_z^2 +\frac{b}{2}\frac{M_z^4}{M_0^2}+a^2 \left(\frac{\partial M_z}{\partial x} \right)^2 \right].
\end{multline}
Here $\tau=(T-\theta)/\theta$ is the reduced temperature, ${\bf M}=M_z(x){\bf z}$ is the magnetization, ${\bf B}=B_z(x){\bf z}$ is the magnetic field with the corresponding vector potential ${\bf A}=A(x){\bf y}$, $\lambda$ is the London penetration depth, $n$ is the concentration of magnetic atoms, $M_0$ is the saturation magnetization at $T=0$, and ${\tilde \theta}=M_0^2/n$. The estimates for the coefficients $a$ and $b$ give $b \sim 1$ and $a \ll \lambda$.

In the cooling regime the first sinusoidal harmonic of magnetization characterized by the wave vector $q_m$ emerges at $T=\theta_m \lesssim \theta$. To calculate the shift $\tau_m=(\theta-\theta_m)/\theta$ one may neglect the term $\propto M_z^4$ in Eq.~(\ref{F_FS}) and make the Fourier expansion: $M_z(x)=\int M_q \exp(iqx)dq/2\pi$.  Then using Maxwell equations we  rewrite the averaged free energy $\bar{F}V=\int F dV$:

\begin{equation}
\label{Fbar_Fourier}
\bar{F}=\sum_q |M_q|^2\left[2\pi /(1+\lambda^2 q^2)+ \Gamma(\tau +a^2q^2)  \right],
\end{equation}
where $\Gamma=n{\tilde \theta}/M_0^2$. The condition ${\bar F}=0$ defines the dependence $\tau(q)$ which minimum corresponds to the actual temperature shift $\tau_m$. The result depends on the value $\lambda$. If $\lambda<a\sqrt{\Gamma/2\pi}$ only the uniform state with  $q_m=0$ should appear while for $\lambda>a\sqrt{\Gamma/2\pi} $ the free energy minimum corresponds to the sinusoidal profile $M_z(x)$ with $q_m=\left(2\pi/\Gamma\lambda^2 a^2\right)^{1/4}$, and we find  $\tau_m=(2a/\lambda)\sqrt{2\pi/\Gamma}$. The period of the emerging magnetic structure is smaller than $\lambda$, which makes this structure compatible with superconductivity due to  the weak Meissner screening.

%%%%%%%%%%%%%%%%%%%%%%%%%%%%%%%%%%%%%%%%%%%%%%%%%%%%%%%%%%%%%%%%%%%%%%%%%%%%%%%%%%%%%%%%%%
\begin{figure}[t!]
\includegraphics[width=0.8\linewidth]{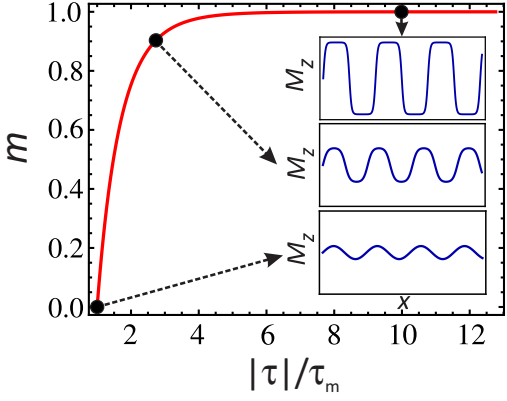}
\caption{The dependence of the parameter $m$ describing the form of magnetic domains on temperature $\tau=(T-\theta)/ \theta$.} \label{Fig:sn_m(tau)}
\end{figure}
%%%%%%%%%%%%%%%%%%%%%%%%%%%%%%%%%%%%%%%%%%%%%%%%%%%%%%%%%%%%%%%%%%%%%%%%%%%%%%%%%%%%%%%%%%

While further cooling below $\theta_m$ the emerging higher harmonics $M_q$ result in the crossover from the sine magnetization profile to the step-like domains $M_z(x)=\pm M$ with the increasing size. The wave vector $Q$ of the domain structure is determined by the balance between the energy of Meissner currents tending to increase $Q$ and the domain walls energy which favors small $Q$ values. In the limit $Q\lambda \gg 1$ the first contribution is proportional to $M^2/(\lambda^2 Q^2)$ while the estimate for the energy of the linear domain walls appearing in the systems with strong magnetic anisotropy gives  $\Gamma a \sqrt{|\tau|}M^2 Q$. The minimization of the resulting free energy shows that the wave vector $Q \sim q_m (\tau_m / |\tau|)^{1/6}$ decreases when cooling the sample.

\begin{figure}[t!]
\includegraphics[width=0.9\linewidth]{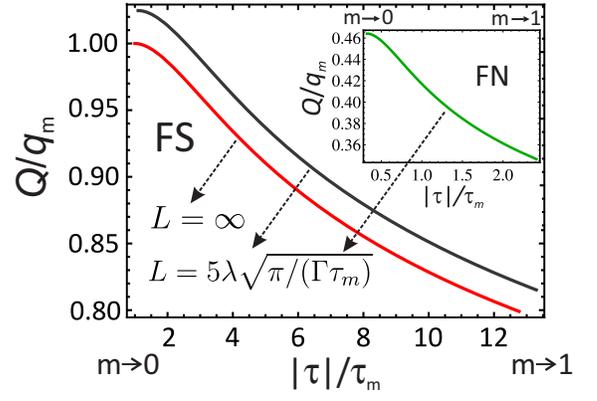}
\caption{The dependence of DS wave vector $Q$ on temperature $\tau=(T-\theta)/ \theta$ for infinite FS (red curve) and FS of the thickness $L$ (black curve). Inset: the same for FN slab.} \label{Fig:sn_Q(tau)}
\end{figure}

The above conclusion is perfectly supported by the accurate calculations. In the easy-axis ferromagnets one can choose the ansatz for the magnetization in the form of the elliptic sine function: $M_z(x)=M {\rm sn} \left[2K(m)Qx / \pi \right]$. Here  $K(m)$ is the elliptic integral and the parameter $m$ controls the shape of $M_z(x)$ profile. Such ansatz perfectly describes the gradual transition between the sine magnetization ($m=0$) and the step-like one ($m \rightarrow 1$). Substituting the Fourier components of $M_z(x)$ into (\ref{Fbar_Fourier}), restoring the term $\propto M_z^4$ and minimizing the resulting functional we obtain the analytical expressions reflecting the temperature evolution of the values $m$, $Q$ and $M$ \cite{supp}. The results confirm the very fast emergence of the well-developed DS below $\theta_m$ (see Fig.{\ref{Fig:sn_m(tau)}}) with the increasing domain size (see Fig. \ref{Fig:sn_Q(tau)}). In the most interesting case of the well-developed DS, i.e. $m \rightarrow 1$, the wave vector $Q$, magnetization $M$ and free energy ${\bar F}$ take the form:
\begin{equation}
\label{M(tau)_domain}
M=\frac{M_0\sqrt{\tau_m}}{\sqrt{b}}  \left[\frac{|\tau|}{\tau_m} -\frac{2^{2/3}}{3}\left(\frac{|\tau|}{\tau_m} \right)^{1/3}\right]^{1/2},
\end{equation}
\begin{equation}
\label{Q(tau)_domain}
Q = q_m\frac{\pi}{2^{4/3}}\left(\frac{\tau_m}{|\tau|} \right)^{1/6},~~~{\bar F}=-\Gamma\frac{b}{2}\frac{M^4}{M_0^2}.
\end{equation}

%%%%%%%%%%%%%%%%%%%%%%%%%%%%%%%%%%%%%%%%%%%%%%%%%%%%%%%%%%%%%%%%%%%%%%%%%%%%%%%%%%%%%%%%%%
%%%%%%%%%%%%%%%%%%%%%%%%%%%%%%%%%%%%%%%%%%%%%%%%%%%%%%%%%%%%%%%%%%%%%%%%%%%%%%%%%%%%%%%%%%

Note that well below $\theta$ the growing magnetization $M(\left|\tau\right|)$ may become large enough to induce Abrikosov vortices. The resulting coexistence phase should emerge through the first order transition. In the presence of the vortex lattice the screening Meissner currents are small and, thus, at temperatures below the transition point instead of DS one should have the uniform magnetization. The thermodynamic critical temperature $T_{M-V}$ of such transition can be obtained from the comparison between the DS and VS free energies. For the well-developed step-like profile $M_z(x)$ the former energy takes the form (\ref{Q(tau)_domain}) while the latter one reads  \cite{DeGennes}
\begin{equation}
\label{F_v}
F_{v}=\frac{(B_z-4\pi M_z)^2}{8\pi}+\Gamma \left(\tau M_z^2+\frac{b}{2}\frac{M_z^4}{M_0^2}\right) +\frac{B_z \tilde{H}_{c1}}{4\pi}.
\end{equation}
Here $\tilde{H}_{c1}=H_{c1}{\rm ln}[\beta d/\xi] /{\rm ln} [\lambda / \xi] $, where $H_{c1}$ is the lower critical field, $\xi$ is the superconducting coherence length, $d^2=2 \Phi_0/ (\sqrt{3}B_z)$,  $\Phi_0$ is the superconducting flux quantum, $\beta=0.381$ is the geometrical factor relevant for the triangular vortex lattice.

In the case $\tilde{H}_{c1} \ll 4\pi M_z$ which is typical for FSs minimizing  (\ref{F_v}) with respect to the magnitude of the uniform magnetization $M_z$ we obtain the free energy of the VS:
\begin{equation}
\label{F_v_min}
F_{v}=-\Gamma M_0^2\tau^2/(2b)+M_0\tilde{H}_{c1}\sqrt{|\tau|/ b}.
\end{equation}
For the reasonable choose of parameters we expect $|\tau_{M-V}|=(\theta - T_{M-V})/ \theta \gg \tau_m$, thus, the temperature of the phase transition between the DS and the VS is
\begin{equation}
|\tau_{M-V}|=\frac{\tau_m}{4}\left(\frac{24\pi \sqrt{b}}{\Gamma} \frac{1}{\tau_m^{3/2}} \frac{\tilde{H}_{c1}}{4\pi M_0}\right)^{6/5}.
\end{equation}

However, the evolution of the magnetic order near the temperature $T_{M-V}$ should reveal hysteresis behavior.  Indeed, in the cooling process the vortices cannot enter the sample at $T_{M-V}$ because of the Bean-Livingston barrier which vanishes only at the temperature $T_{cm}<T_{M-V}$. The profile of this barrier $U(x)$ is determined by the interplay between the energies of the vortex interaction with Meissner currents and with the antivortex located in the neighbouring domain. For the step-like profile of the magnetization the profile $U(x)$ has the form
\begin{equation}
U(x)=-\frac{\Phi_0}{4\pi}\int_0^{x}\frac{dB_v(2\tilde{x})}{d\tilde{x}} d\tilde{x}-\frac{4\Phi_0M}{\pi\lambda^2Q^2} \sum_{n=0}^{\infty} \frac{\sin[Qx(2n+1)]}{(2n+1)^3},
\end{equation}
where $B_v$ is the magnetic field produced by the antivortex. 
%Note that in our situation the vortex interacts with a real anti-vortex in the neighboring domain in contrast with the superconductor/vacuum interface where the vortex is attracted by its image. 
The condition of the barrier vanishing $(dU /dx)|_{x \rightarrow \xi}=0$ allows us to calculate the value $\tau_{cm}=(T_{cm}-\theta)/ \theta$ using (\ref{Q(tau)_domain}) and (\ref{M(tau)_domain}):
\begin{equation}
|\tau_{cm}|=\frac{[\sqrt{1+9\gamma}+1]^{3/2}}{\sqrt{54}}\tau_m,  ~~\gamma=\frac{4\pi b}{\Gamma \tau_m^2} \left(\frac{H_{cm}}{4\pi M_0} \right)^2,
\end{equation}
where $H_{cm}$ is the thermodynamic critical field. 

Taking the parameters of EuFe$_2$As$_2$ \cite{Nandi} we may estimate the ratio between $T_{M-V}$ and $T_{cm}$. Since $\tau_m$ is rather small $\tau_m \sim 10^{-3} \ll 1$, we have $\gamma \gg 1$ and $|\tau_{cm}| \approx (\gamma^{3/4}/\sqrt{2})\tau_m$ and, thus, $T_{M-V}$ significantly exceeds $T_{cm}$. However, the calculated value of $T_{cm}$ may be substantially smaller than the temperature of the vortex entry in experiment. Indeed, the presence of accidental vortices trapped, e. g., at the defects, increases $T_{cm}$ since the Meissner currents near the vortex cores are of the order of depairing current which favors the formation the additional vortex-antivortex pairs. Thus, in real type-II superconductors we expect $T_{cm} \lesssim T_{M-V}$. 

%%%%%%%%%%%%%%%%%%%%%%%%%%%%%%%%%%%%%%%%%%%%%%%%%%%%
\begin{figure}[b!]
\includegraphics[width=1\linewidth]{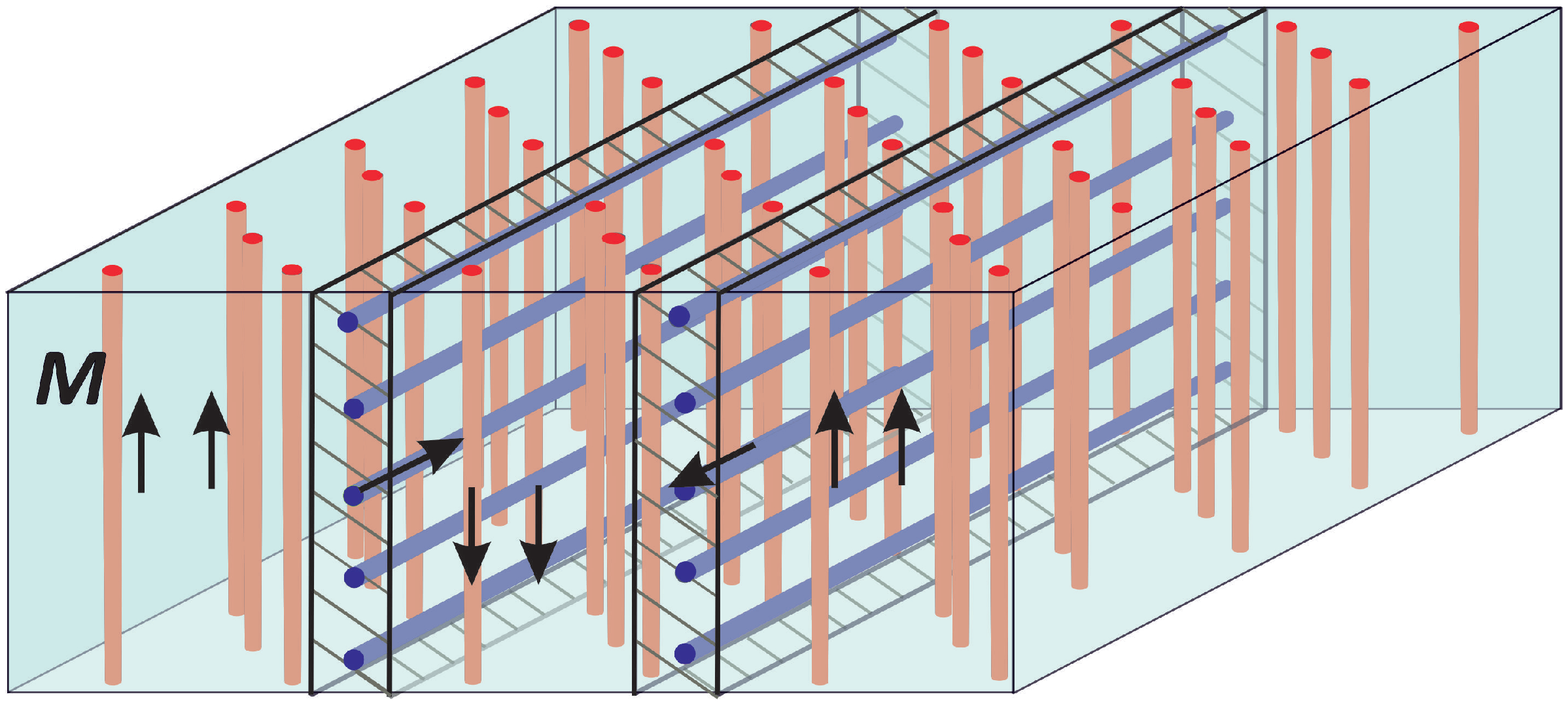}
\caption{``Perpendicular'' vortices appearing at the domain walls of the Bloch type. } \label{Fig_PDW}
\end{figure}
%%%%%%%%%%%%%%%%%%%%%%%%%%%%%%%%%%%%%%%%%%%%%%%%%%%%

Up to now we have not accounted the finite size of the sample. For the slab of the finite thickness $L$ the domain structure produces the stray magnetic field which decays at distances $\propto Q^{-1}$ from the slab surface. The corresponding contribution to the free energy tends to shrink the domains, thus, making the vector $Q$  higher for the thinner samples at a fixed $T$ (see Fig.~\ref{Fig:sn_Q(tau)}) \cite{supp}.  However, even for the thin-film FSs the stray field effect does not qualitatively change the dependence $Q(\tau)$ and the domain size remains increasing upon cooling the sample.

Interestingly, in the finite samples the emergence of the dense vortex lattice below $T_{cm}$ on top of magnetic DS should result in the substantial growth in the domain size. In this coexistence phase the Meissner screening is almost destroyed and the situation is fully analogous to the non-superconducting ferromagnetic film (FN). A such a film the ferromagnetic domain structure should appear to minimize the stray field.	Calculating the dependence $Q(\tau)$ for this case we find that at a given $T$ the domains is significantly larger if the superconductivity is destroyed (see the inset in Fig.~\ref{Fig:sn_Q(tau)}). Such increase of the domain size associated with the transition to the VS was clearly observed in \cite{Veshchunov, Vinnikov}. Note that the similar effect should exist in FSs with $T_c < \theta$ \cite{Faure, Dao, Khaymovich}.

On experiment (see supplementary materials to Ref.[\onlinecite{Vinnikov}]) the observed in  EuFe$_2$As$_2$ low temperature domain structure is very similar to the branched domain patterns in normal ferromagnets (see for example section 5.2 in  Ref.[\onlinecite{Hubert}]). The dense vortex structure in ferromagnetic domains in  EuFe$_2$As$_2$ make them equivalent to that in the normal ferromagnets. The recent observation of the low temperature  domain structure in the overdoped normal EuFe$_2$As$_2$ show that it is basically the same as in the optimally doped superconducting EuFe$_2$As$_2$  \cite{Stolyarov}.

The reason of the branching of the domains in the ferromagnets is related with the fact that the stray field energy increases with the increase of the domain size faster than the domains wall energy and starting some critical thickness the branching of the domain becomes energetically favorable \cite{LL}.
At low temperature the internal magnetic field in EuFe$_2$As$_2$ is rather large $B(0) \approx 9$kOe (which strongly exceeds the low critical field) \cite{Vinnikov} and the period of the Abrikosov lattice is of the order of 40nm. So it is much smaller than characteristic size of the domain pattern and then the vortices simply decorate without changing the usual mechanism of the formation of ria-cost magnetic domain structure observed in EuFe$_2$As$_2$ \cite{Vinnikov} at low temperature. Note that the calculation of the branched domain patterns in the ferromagnets is beyond the scope of our paper.

Interestingly, in the VS near the domain walls the vortices can become oriented perpendicular to the vortices in domains (see Fig.~\ref{Fig_PDW}).  This phenomenon originates from the transformation of the linear domain walls with only one magnetization component $M_z$ to the Bloch type domain walls having an additional component $M_y$. Exactly this component favors the vortex core directed along the $y$-axis near the domain walls. The comparison between the free energies \cite{supp} shows that for EuFe$_2$As$_2$ the linear domain walls exists in the temperature interval $\Delta T \sim 0.3$K near $\theta$ while for lower $T$ the Bloch domain walls appear. The subsequent emergence of the perpendicular vortices is favorable if $4\pi M \ge 4\pi M_{min}= (2\lambda/\pi\delta)H_{c1}$, where $\delta$ is the domain wall width \cite{supp}. For EuFe$_2$As$_2$ the estimates give $4\pi M_{min} \sim 0.4$kOe while $4\pi M \sim  10^4$Oe \cite{Vinnikov, Nandi}. Thus, the condition of 
perpendicular vortices appearance is fulfilled.

To sum up, we have described the temperature evolution of the magnetic domain structures and vortex-domain coexistence in ferromagnetic superconductors with purely electromagnetic interaction. The developed theory not only describes the transition from the short-period Meissner domain phase to the vortex phase recently observed in EuFe$_2$As$_2$ but also predicts the formation of "perpendicular vortices" at the domain walls.

\acknowledgements

The authors thank I. Veshchunov, L. Ya. Vinnikov, A. S. Mel'nikov and I. A. Shereshevskii for fruitful discussions. This work was supported by the French ANR SUPERTRONICS and OPTOFLUXONICS, EU COST CA16218 Nanocohybri,  Russian Science Foundation under Grant No. 15-12-10020, Foundation for the advancement of theoretical physics "BASIS", Russian Presidential Scholarship SP-3938.2018.5 and Russian Foundation for Basic Research under Grant No. 18-02-00390.

\end{document}